\documentclass{desyproc}

\begin{document}
\title{Search for a $\pi\Lambda N-\pi\Sigma N$ dibaryon in p+p@3.5~GeV}

\author{{\slshape J.C.~Berger-Chen$^{1,2}$, L.~Fabbietti$^{1,2}$} for the HADES Collaboration\\[1ex]
$^1$Physik Department E12, Technische Universit\"{a}t M\"{u}nchen, 85748~Garching, Germany\\
$^2$Excellence Cluster 'Origin and Structure of the Universe', 85748~Garching, Germany}

\contribID{101}

\confID{8648}  
\desyproc{DESY-PROC-2014-04}
\acronym{PANIC14} 
\doi  

\maketitle

\begin{abstract}
This work is dedicated to the search for a $\pi\Lambda N\!-\!\pi\Sigma N$ resonance $\mathcal{Y}$ with the quantum numbers ($Y,I,J^P$) = (1,$\frac{3}{2}$,$2^+$). The double charged $\Sigma(1385)N\!-\!\Delta(1232)Y$ quasibound state was looked for in the reaction $pp\rightarrow \mathcal{Y}^{++}K^0$ with its unique decay into $\Sigma^+$ and proton measured with the HADES setup at a kinetic beam energy of 3.5~GeV. The analysis including background determination and a description of the data with a $K^0_S$ Monte Carlo cocktail are presented.
\end{abstract}

\section{Introduction}
Recently, a relativistic three-body Fadeev formalism suitable for two-body p-wave interactions was applied to calculate the $\pi\Lambda N\!-\!\pi\Sigma N$ coupled channel system with $I=\frac{3}{2}$ and $J^P=2^+$ \cite{Garcilazo:2012rh}. Dominant p-wave interactions in the $\pi N$ and $\pi\Lambda\!-\!\pi\Sigma$ channels were found with a large contribution of the $\Delta(1232)$ and $\Sigma(1385)$ resonances. A rather robust $\pi\Lambda N$ resonance located around 10-20~MeV/c$^2$ below the $\pi\Sigma N$ threshold was obtained, which can be viewed as an s-wave dibaryon $\mathcal{Y}$ with ($Y,I,J^P$) = (1,$\frac{3}{2}$,$2^+$) equivalent to an $\Sigma(1385)N\!-\!\Delta(1232)Y$ quasibound state bound by over 50~MeV. The double charged state can be uniquely measured in p+p collisions at energies above the $\Sigma(1385)$ production threshold ($pp\!\rightarrow\!\mathcal{Y}^{++}K^0\!\rightarrow\!\Sigma^+pK^0$), whereas the $\mathcal{Y}^+$ formed together with a $K^+$ ($\mathcal{Y}^+\!\rightarrow\!\Sigma^+n/\Sigma^0p$) may 
not be distinguishable from the decay of a $\bar{K}\!N\!N$ quasibound state ($K^-pp\!\rightarrow\!\Sigma^+n/\Sigma^0p$). Thus, the HADES p+p data measured at 3.5~GeV, which is well above $\Sigma(1385)$ production threshold \cite{Agakishiev:2011qw}, are perfectly suited to search for the $\mathcal{Y}^{++}$ dibaryon in its unique decay. Thereby, knowledge gained by a previous analysis \cite{Agakishiev:2014nim} on exclusive $K^0$ production channels in the same data set were incorporated meaning that the determined cross sections and angular anisotropies were included in this work. The main result of that analysis was the finding of predominant contributions by $K^0$ channels associated with resonances ($\Delta(1232)$ and $\Sigma(1385)$) thus confirming earlier observations made in the study of resonance contributions ($\Lambda(1405)$, $\Sigma(1385)^0$, $\Lambda(1520)$, $\Delta(1232)$, $N^*$, $K^{*0}$) to the $\Sigma^{\pm}\pi^{\pm}pK^{+}$ final states \cite{Agakishiev:2012qx}.

\section{The HADES experiment}
The \textbf{H}igh-\textbf{A}cceptance \textbf{D}i-\textbf{E}lectron \textbf{S}pectrometer (HADES) \cite{Agakishiev:2009am} is a versatile detector system located at GSI Helmholtzzentrum (Darmstadt, Germany) and is provided by the SIS18 with heavy ion beams of 1-2~AGeV or proton beams up to 3.5~GeV impinging on a fixed target. The HADES setup has an azimuthal coverage of 85\% and an acceptance in polar angles from 18$^\circ$ to 85$^\circ$. The momentum resolution was determined to be $\Delta p/p\approx 3\%$.

The most important detector components used in the presented analysis are first of all the Multi-Wire Drift Chambers (MDCs), where two layers are mounted in front of the superconducting magnet with a toroidal field and two layers behind it to help in track finding, momentum reconstruction and particle identification via the specific energy loss information. Furthermore, a Time-Of-Flight wall at the end of the setup allows for online multiplicity triggering to enhance inelastic events in the recorded data sample.
\begin{figure}[h]
\centering
\begin{minipage}[t]{0.48\textwidth}
\centering
\includegraphics[height=0.9\textwidth, angle=90]{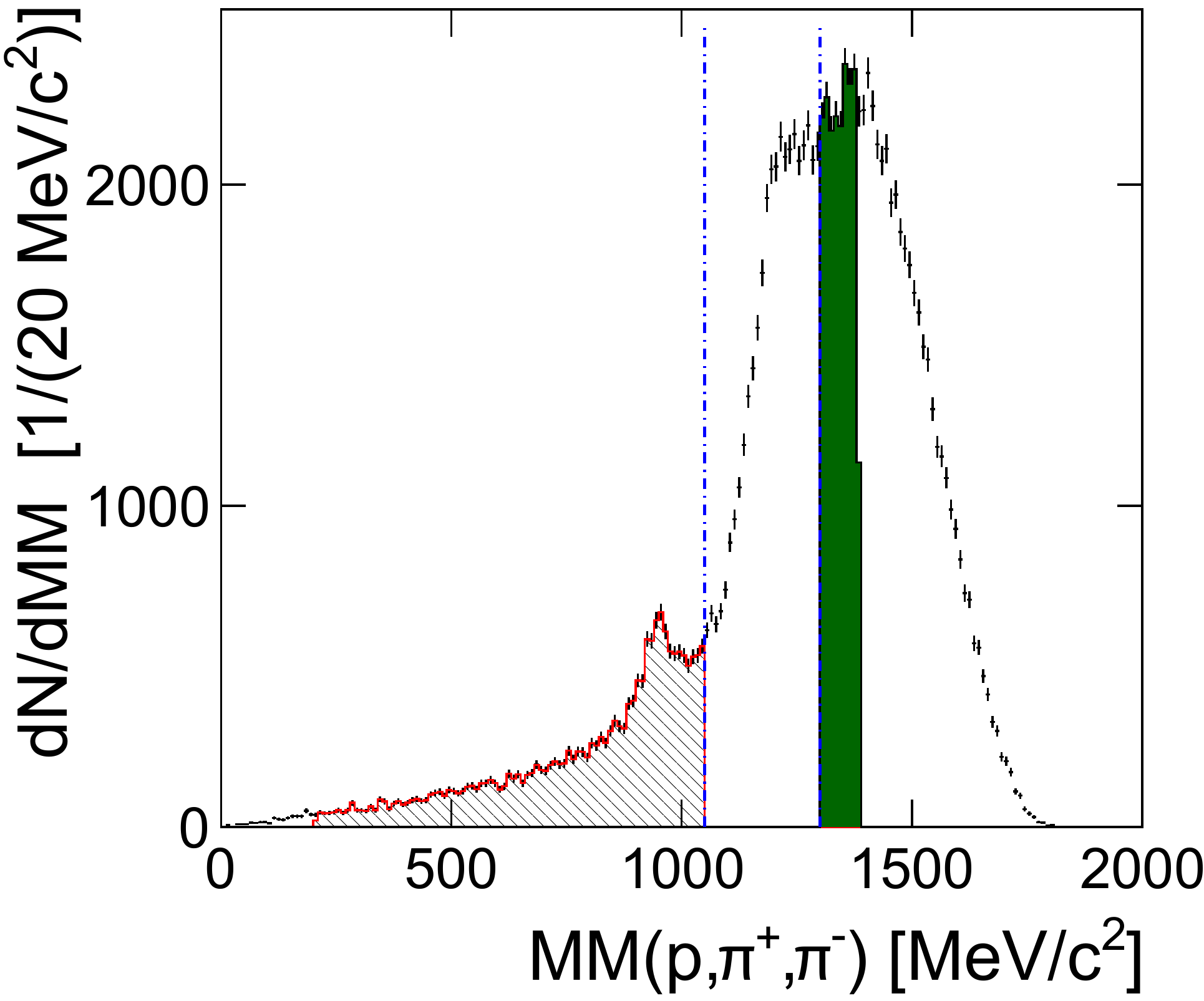}
\caption{Missing mass spectrum with respect to the proton, $\pi^+$ and $\pi^-$. The red-hatched and green-filled areas indicate the mass regions used for the low mass (LM) and the high mass (HM) sideband sample. The blue dash-dotted lines show the cut on the $\Sigma^+$ mass.}\label{Fig:MMpK0S}
\end{minipage}
\hspace{0.02\textwidth}
\begin{minipage}[t]{0.48\textwidth}
\centering
\includegraphics[height=0.9\textwidth, angle=90]{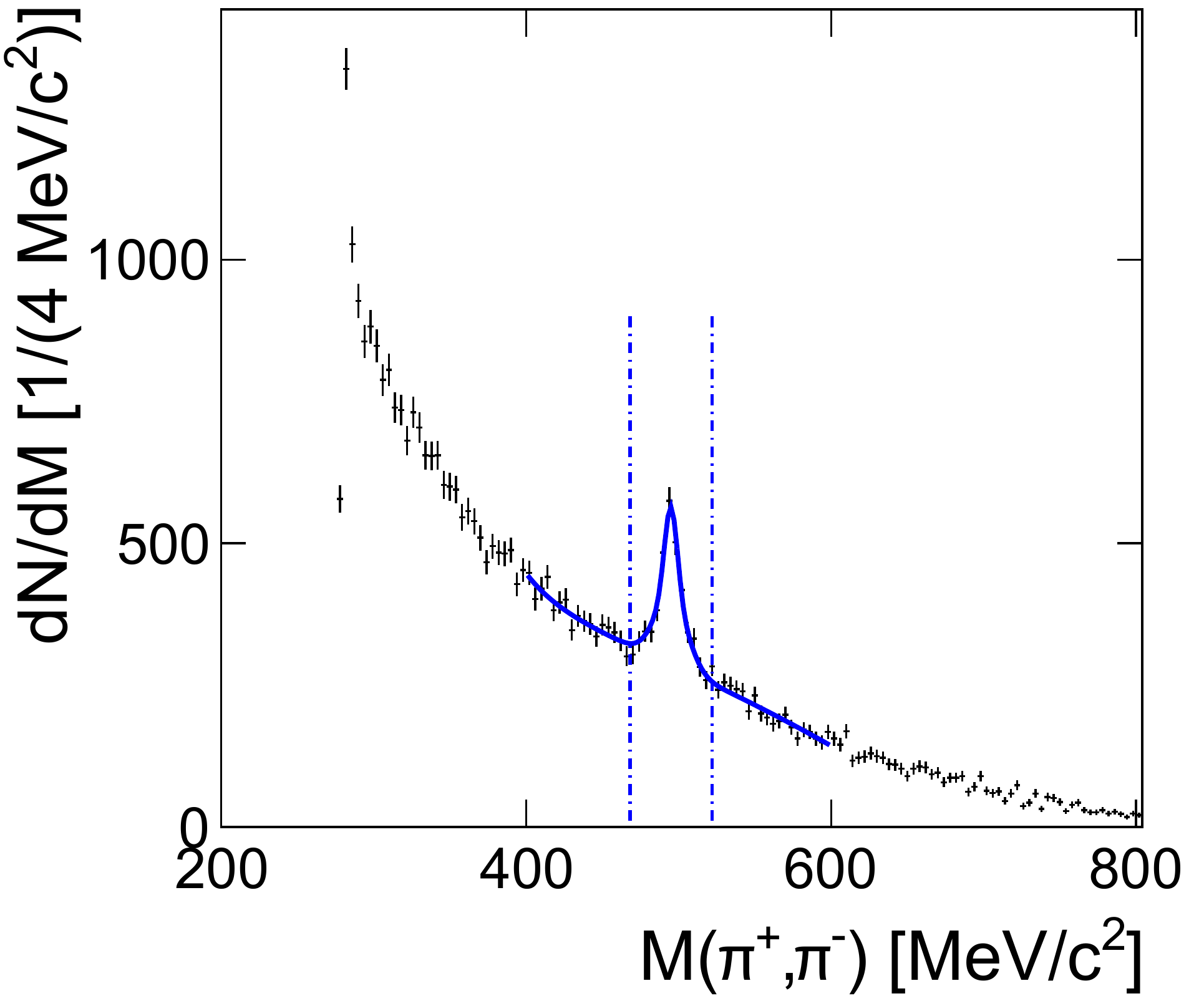}
\caption{$\pi^+\pi^-$-invariant mass spectrum including secondary vertex cuts \cite{Agakishiev:2014nim} and fitted with the sum of two Gaussians, a Landau and a polynomial. The blue dash-dotted lines indicate the 3$\sigma$-cut region on the $K^0_S$ mass.}\label{Fig:IMK0S}
\end{minipage}
\end{figure}

\section{The analysis and preliminary results}
As mentioned above the $\pi\Lambda N\!-\!\pi\Sigma N$ resonance, which can be denoted as $\mathcal{Y}$, can be uniquely measured through the reaction $pp\!\rightarrow\!\mathcal{Y}^{++}K^0\!\rightarrow\!\Sigma^+pK^0$. Since the long-lived $K^0_L$ cannot be detected by the HADES setup, this analysis concentrates on the short-lived $K^0_S$ and its charged decay into $\pi^+$ and $\pi^-$ ($BR\!\approx\!69.20$\%). A four particle selection was implemented to enhance the above reaction by choosing events with a proton, $\pi^+$, $\pi^+$ and a $\pi^-$. The proton is thereby assumed to be produced directly, whereas one of the $\pi^+$ should originate from the decay of the $\Sigma^+$. This way, only the decay of the $\Sigma^+$ into a neutron and a $\pi^+$ ($BR\!\approx\!48.31$\%) was considered neglecting the branching into proton and $\pi^0$  ($BR\!\approx\!51.57$\%), since the heavier baryon is more likely to be boosted into the forward region, where HADES has no acceptance.

\subsection{Background determination}
\begin{figure}[h]
\centering
\begin{minipage}[c]{0.73\textwidth}
\includegraphics[width=0.73\textwidth, angle=90]{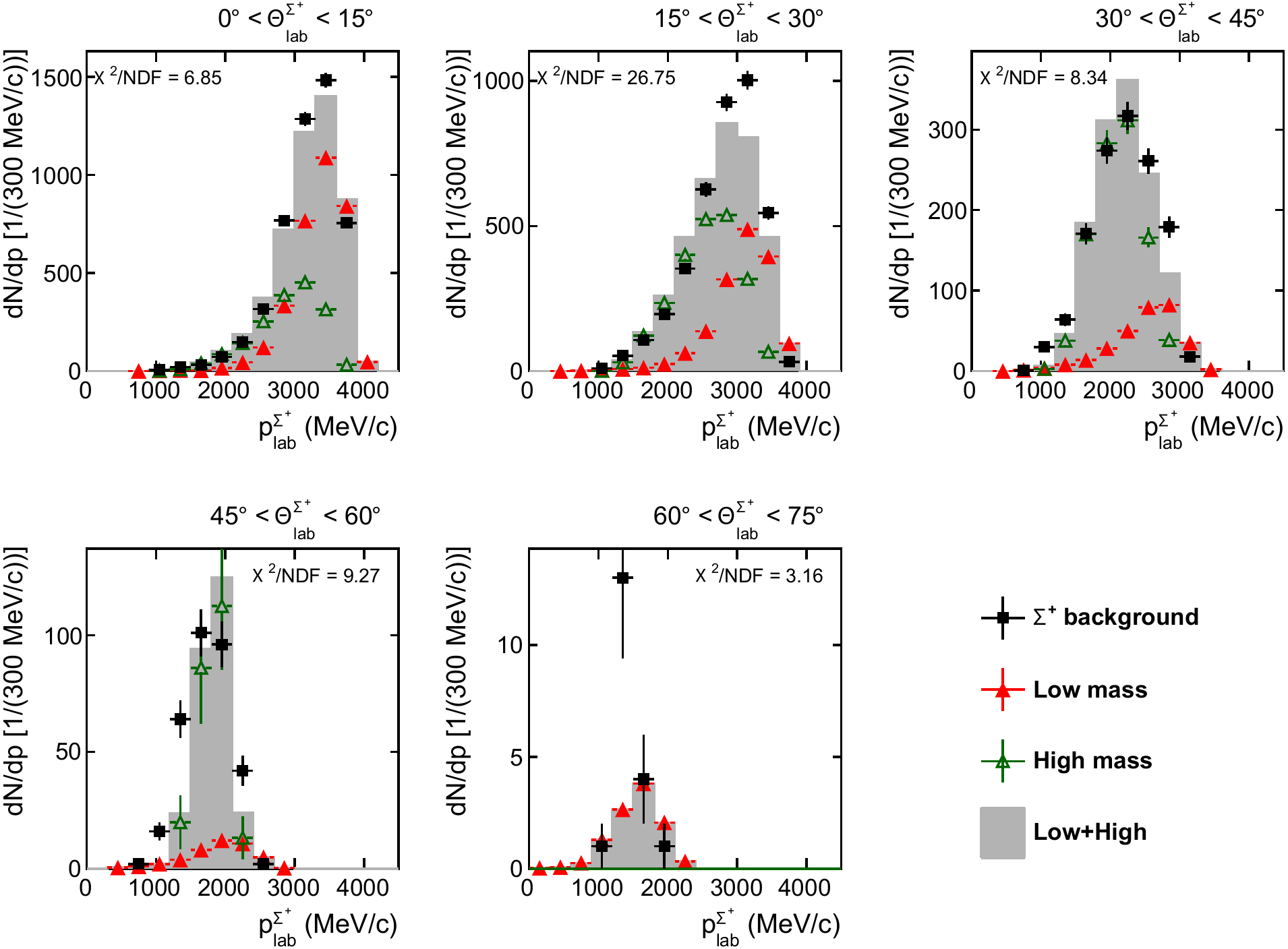}
\end{minipage}
\begin{minipage}[c]{0.25\textwidth}
\caption{Momentum distribution of the $\Sigma^+$ background in five bins of $\Theta_{lab}^{\Sigma^+}$ with the cuts: MM(p,$\pi^+$,$\pi^-$) = 1050-1300~MeV/c$^2$ and 810~MeV/c$^2$ $>$ MM(p,$\pi^+$,$\pi^+$,$\pi^-$) $>$ 1045~MeV/c$^2$. Overlayed are the spectra for the LM and HM sideband samples with their sum in gray.}\label{Fig:Sideband}
\end{minipage}
\end{figure}
The background determination in the four particle data sample is a crucial step in this analysis and consists mainly of combinatorial background generated by non-strange reactions. Contributions from other $K^0_S$ production channels can be modeled in a first attempt by an incoherent Monte Carlo cocktail, which individual cross sections were obtained in a separate analysis of the same data set \cite{Agakishiev:2014nim}. To emulate the background, a so-called sideband analysis was performed on the missing mass distribution to the proton, $\pi^+$ and $\pi^-$ (Fig. \ref{Fig:MMpK0S}). Only a rough cut on the primary vertex was applied here to reject off-target events. Furthermore, no additional mass cut, e.g. on the $K^0_S$, was applied to keep enough statistics for the sideband analysis. A data sample was defined in this spectrum, which contains data with missing masses lower (LM) and higher (HM) than the $\Sigma^+$ (1050~MeV/c$^2$ $>$ MM(p,$\pi^+$,$\pi^-$) $>$ 1300~MeV/c$^2$) and thus include only background. They are indicated as red-hatched and green-filled areas in the plot and are chosen such to have about the same integral. This background sample needs to reproduce the kinematics, which remains in the $\Sigma^+$ mass region. Therefore, the momentum distributions of the sideband samples LM and HM were fitted in five individual $\Theta_{lab}$ bins to describe the momentum distribution of a $\Sigma^+$ background sample, which was obtained through a cut on the missing mass spectrum to the four particles proton, $\pi^+$, $\pi^+$ and $\pi^-$ (810~MeV/c$^2$ $>$ MM(p,$\pi^+$,$\pi^+$,$\pi^-$) $>$ 1045~MeV/c$^2$). From the exclusive analysis discussed in \cite{Agakishiev:2014nim} it is known that the reaction $pp\!\rightarrow\!\Sigma^+pK^0_S$ does not contribute in these mass regimes. Figure \ref{Fig:Sideband} shows the mentioned momentum distributions of $\Sigma^+$ candidates and quotes the achieved $\chi^2/NDF$ for each $\Theta_{lab}$ bin, which mostly stays below 10 except for $\Theta_{lab} = 15^\circ-30^\circ$ with $\chi^2/NDF = 26.75$. The 
conclusion is, that the sideband sample does not fully describe the $\Sigma^+$ background and further work is necessary here. However, the description is good enough to use the so determined relative contributions of the LM and HM samples to the full background model for a first try to search for a signal of the $\mathcal{Y}^{++}$ dibaryon, which should be located around 10-20~MeV/c$^2$ below the $\pi\Sigma N$ threshold (2267~MeV/c$^2$) \cite{Garcilazo:2012rh}.
\begin{figure}[h]
\centering
\begin{minipage}[t]{0.48\textwidth}
\centering
\includegraphics[height=0.9\textwidth, angle=90]{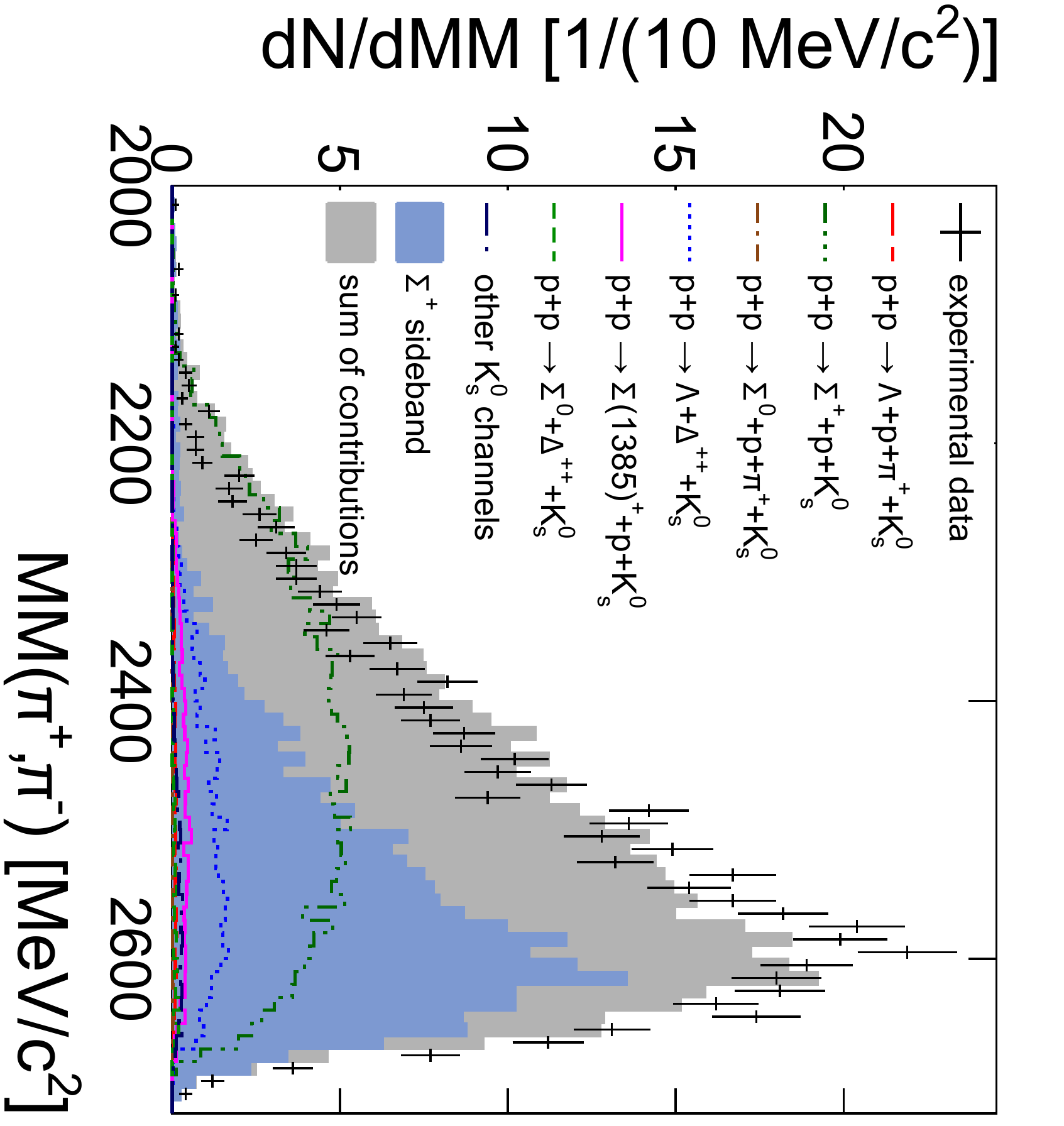}
\caption{Missing mass spectrum with respect to the $\pi^+$ and $\pi^-$ with a cut on the $K^0_S$ and the $\Sigma^+$ mass. The contributions of the $K^0_S$ reactions are according to the findings of \cite{Agakishiev:2014nim}.}\label{Fig:MMK0S}
\end{minipage}
\hspace{0.02\textwidth}
\begin{minipage}[t]{0.48\textwidth}
\centering
\includegraphics[height=0.9\textwidth, angle=90]{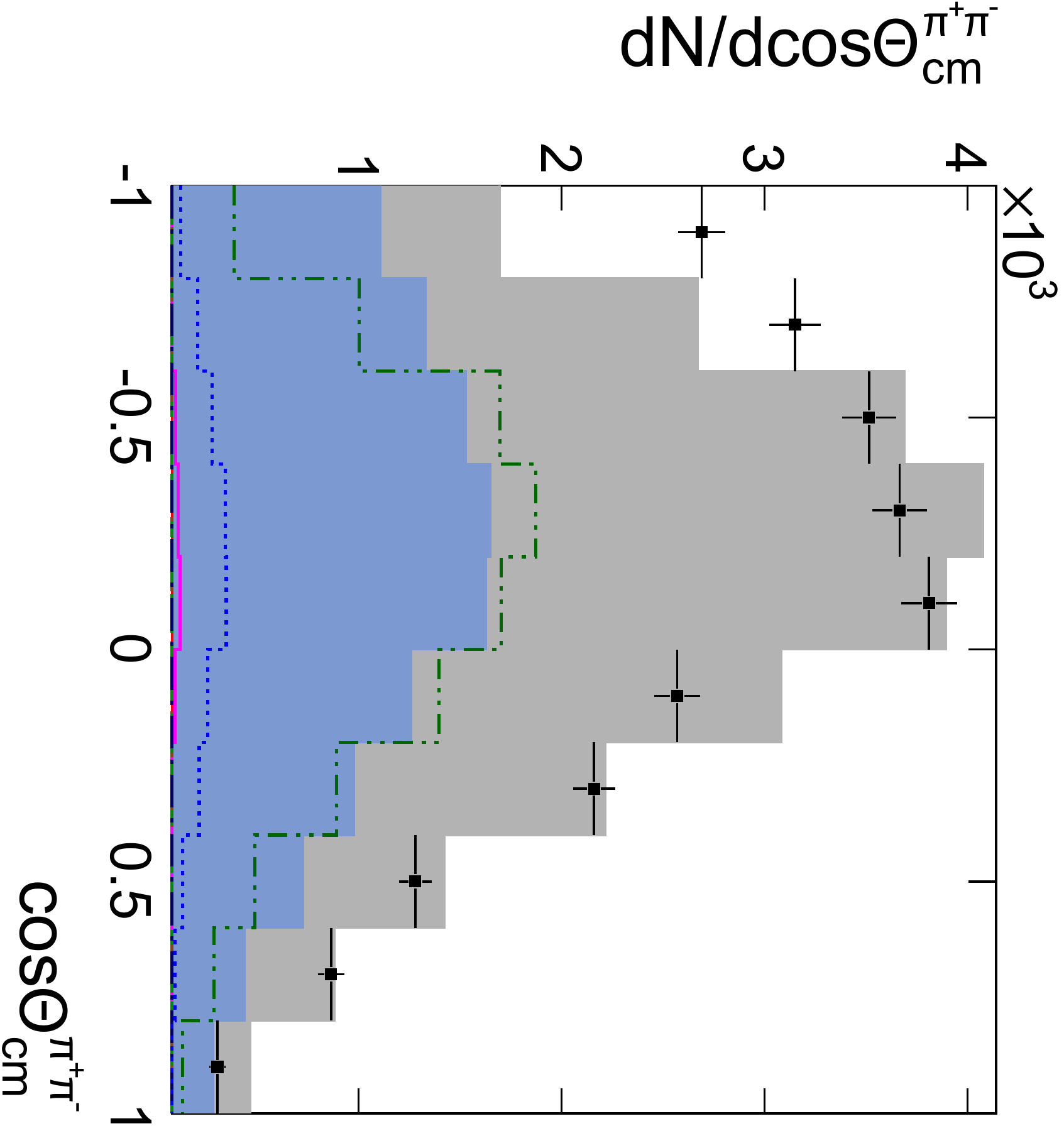}
\caption{Angular distribution of the $\pi^+\pi^-$-system in the center-of-mass with a cut on the $K^0_S$ and on the $\Sigma^+$ mass. The color code is the same as written in the label of Figure \ref{Fig:MMK0S}.}\label{Fig:CosThcmK0S}
\end{minipage}
\end{figure}

\subsection{Preliminary results}
In Figure \ref{Fig:MMK0S} the MM($\pi^+,\pi^-$) distribution is depicted, which contains a cut on the $\Sigma^+$ mass (1050~MeV/c$^2$ $<$ MM(p,$\pi^+$,$\pi^-$) $<$ 1300~MeV/c$^2$) and a 3$\sigma$-cut on the $K^0_S$ mass determined in the $\pi^+\pi^-$-invariant mass distribution (Fig. \ref{Fig:IMK0S}) by a fit with a sum of two Gaussians for the signal and a Landau plus a polynomial function for the background. However, no enhancement is seen in the interesting region below the $\pi\Sigma N$ threshold. The missing mass spectrum is rather well described by the incoherent $K^0_S$ cocktail determined in \cite{Agakishiev:2014nim} and the background model, from which the relative contribution was obtained by a simultaneous fit to this missing mass spectrum and the $\pi^+\pi^-$-angular distribution $cos\Theta_{cm}^{\pi^+\pi^-}$ (Fig. \ref{Fig:CosThcmK0S}) with fixed cross sections and angular anisotropies for the $K^0_S$ reactions. Although the $\chi^2/NDF$ extracted from the simultaneous fit to the two spectra of 
5.07 is rather good, some disagreement between model and data is observed at $cos\Theta_{cm}^{\pi^+\pi^-} < -0.5$, which can be due to the imperfect background model.

\section{Summary and outlook}
The $\pi\Lambda N\!-\!\pi\Sigma N$ resonance $\mathcal{Y}$ with ($Y,I,J^P$) = (1,$\frac{3}{2}$,$2^+$) was searched for in the reaction $pp\!\rightarrow\!\mathcal{Y}^{++}K^0\!\rightarrow\!\Sigma^+pK^0\!\rightarrow\!n\pi^+pK^0$. Thereby, a sideband analysis was performed to model the $\Sigma^+$ background in the distribution MM($p,\pi^+,\pi^-$). A satisfactory description was achieved to use this background model in the further study of the missing mass spectrum MM($\pi^+,\pi^-$) with help of the $K^0_S$ reaction cocktail determined in \cite{Agakishiev:2014nim}. No obvious enhancement is seen in the MM($\pi^+,\pi^-$) below the $\pi\Sigma N$ threshold, in fact the data is described rather well by the model. However, no interference effects were taken into account, which can be only studied with help of a Partial Wave Analysis (PWA) similar as for the search of the $K^-pp$ quasibound state in the same data set \cite{Epple:2014}.

\section{Acknowledgements}
\sloppy
The authors gratefully acknowledge the support of the TUM Graduate School's Faculty Graduate Center Physics at Technische Universit\"at M\"unchen, Germany, 
PTDC/FIS/113339/2009 LIP Coimbra, NCN grant 2013/10/M/ST2/00042 SIP JUC Cracow, Helmholtz Alliance HA216/EMMI GSI Darmstadt, VH-NG-823, Helmholtz Alliance HA216/EMMI TU Darmstadt, 283286, 05P12CRGHE HZDR Dresden, Helmholtz Alliance HA216/EMMI, HIC for FAIR (LOEWE), GSI F\&E Goethe-University, Frankfurt VH-NG-330, BMBF 06MT7180 TU M\"unchen, Garching BMBF:05P12RGGHM JLU Giessen, Giessen UCY/3411-23100, University Cyprus CNRS/IN2P3, IPN Orsay, Orsay MSMT LG 12007, AS CR M100481202, GACR 13-06759S NPI AS CR, Rez, EU Contract No. HP3-283286.


\begin{footnotesize}


\bibliographystyle{unsrt}
\bibliography{berger-chen_jia-chii}

\begin{thebibliography}{1}

\bibitem{Garcilazo:2012rh}
H.~Garcilazo and A.~Gal.
\newblock {Relativistic three-body calculations of a Y=1, I=3/2, JP=2+
  $\pi\Lambda N$ -- $\pi\Sigma N$ dibaryon}.
\newblock {\em Nucl. Phys.}, A897:167--178, 2013.

\bibitem{Agakishiev:2011qw}
G.~Agakishiev et~al.
\newblock {Baryonic resonances close to the $\bar{K}$-N threshold: The Case of
  $\Sigma(1385)^+$ in pp collisions}.
\newblock {\em Phys. Rev.}, C85:035203, 2012.

\bibitem{Agakishiev:2014nim}
G.~Agakishiev et~al.
\newblock {Associate $K^0$ production in p+p collisions at 3.5 GeV: The role of
  $\Delta(1232)^{++}$}.
\newblock {\em Phys. Rev.}, C90:015202, 2014.

\bibitem{Agakishiev:2012qx}
G.~Agakishiev et~al.
\newblock {Production of $\Sigma^{\pm}\pi^{\pm}pK^{+}$ in p+p reactions at 3.5
  GeV beam energy}.
\newblock {\em Nucl. Phys.}, A881:178--186, 2012.

\bibitem{Agakishiev:2009am}
G.~Agakishiev et~al.
\newblock {The High-Acceptance Dielectron Spectrometer HADES}.
\newblock {\em Eur. Phys. J.}, A41:243--277, 2009.

\bibitem{Epple:2014}
E.~Epple.
\newblock {Experimental news from a theoretical state: The
  '\textit{ppK}$\mathbf{^{-}}$'}.
\newblock {\em PoS}, BORMIO2014:049, 2014.

\end{thebibliography}
%

\end{footnotesize}


\end{document}